\documentclass[11pt,a4paper]{article}
\usepackage[utf8]{inputenc}

\usepackage{amssymb}
\usepackage{amsmath}
\usepackage{mathrsfs}
\usepackage{graphicx}
\usepackage{url}
\usepackage{lineno}

\usepackage[round]{natbib}

\title{Stability of Underdominant Genetic Polymorphisms in Population Networks}

\author{\'Aki J.~L\'aruson\footnote{Department of Biology, University of Hawai`i at M\=anoa, Honolulu, Hawai`i 96822}~ and Floyd A.~Reed\footnotemark[1] \footnote{corresponding author: floydr@hawaii.edu}}

\date{\today}

\begin{document}

\maketitle


\begin{abstract}
Heterozygote disadvantage is potentially a potent driver of population genetic divergence.
Also referred to as underdominance, this phenomena describes a situation where a genetic heterozygote has a lower overall fitness than either homozygote.
Attention so far has mostly been given to underdominance within a single population and the maintenance of genetic differences between two populations exchanging migrants.
Here we explore the dynamics of an underdominant system in a network of multiple discrete, yet interconnected, populations. 
Stability of genetic differences in response to increases in migration in various topological networks is assessed.
The network topology can have a dominant and occasionally non-intuitive influence on the genetic stability of the system. 
\end{abstract}

\begin{center}
\end{center}

\section{\label{Introduction} Introduction}

Variation in the fitness of genotypes resulting from combinations of two alleles (\textit{e.g.},  A- and B-type alleles combined into AA-, AB-, or BB-genotypes resulting in $w_{AA}$, $w_{AB}$, and $w_{BB}$ fitnesses respectively) result in different evolutionary dynamics.  
The case in which a heterozygote has a lower fitness than either homozygote, $w_{AB} < w_{AA}$ and $w_{AB} < w_{BB}$, is termed underdominance or heterozygote disadvantage.  In this case 
there is an internal unstable equilibrium so that the fixation or loss of an allele depends on its starting frequency.  In a single population, stable polymorphism is not possible.  However, when certain conditions are met, populations that are coupled by migration (the exchange of some fraction of alleles each generation) can result in a stable selection-migration equilibrium.  This selection-migration equilibrium is associated with a critical migration rate ($m^*$); above this point the mixing between populations is sufficiently high for the system to behave as a single population and all internal stability is lost \citep{Altrock2010}. 

Underdominance can be thought of as an evolutionary bistable switch. From the perspective of game-theory dynamics it can be interpreted as a coordination game \citep{Traulsen2012}.  The properties of underdominance in single and multiple populations have led to proposals of a role of underdominance in producing barriers to gene flow during speciation \citep{Faria2010,Harewood2010} as well as proposals to utilize underdominace both to transform the properties of wild populations in genetic pest management applications \citep{Curtis1968,Davis2001,Sinkins2006,Reeves2014b} and to engineer barriers to gene flow (transgene mitigation) from genetically modified crops to unmodified relatives \citep{Soboleva2003,Reeves2014a}.  

The properties of underdominance in a single population are well understood \citep{Fisher1922,Haldane1927,Wright1931} and the two-population case has been studied in some detail \citep{Karlin1972a,Karlin1972b,Lande1985,Wilson1986,Spirito1991,Altrock2010,Altrock2011}, with fewer analytic treatments of three or more populations  \citep{Karlin1972a,Karlin1972b}.  Simulation-based studies have been conducted for populations connected in a lattice \citep{Schierup1996,Payne2007,Eppstein2009,Landguth2015} and ``wave'' approximations have been used to study the flow of underdominant alleles under conditions of isolation by distance  \citep{Fisher1937,Pialek1997,Soboleva2003,Barton2011}.  Despite these properties and potential applications, underdominance has been relatively neglected in population genetic research \citep{Bengtsson1976}.
A large focus of earlier theoretical work with underdominance was on how new rare mutations resulting in underdominance might become established in a population \citep{Wright1941,Bengtsson1976,Hedrick1981,Walsh1982,
Hedrick1984,Lande1984,Lande1985,Barton1991,Spirito1992}. However, here we are addressing the properties of how underdominant polymorphisms may \emph{persist} once established within a set of populations rather than how they were \emph{established} in the first place.   
  
   We explicitly focus on discrete populations that are connected by migration in a population network. We have found that the topology of the network has a profound influence on the stability of underdominant polymorphisms that has been otherwise overlooked.  This influence is not always intuitive \textit{a priori}.  These results have implications for the effects of network topology on a dynamic system \citep[see for review][]{Strogatz2001}, particularly for interactions related to the coordination game \citep[such as the stag hunt,][]{Skyrms2001}, theories of speciation, the maintenance of biological diversity, and applications of underdominance to both protect wild populations from genetic modification or to genetically engineer the properties of wild populations---depending on the goals of the application.  

\section{\label{Methods and Results} Methods and Results}

We are considering simple graphs in the sense of graph theory to represent the population network: each pair of nodes can be connected by at most a single undirected edge. A graph $\mathbb{G}=\left( \mathscr{N} , \mathscr{E} \right)$, is constructed from a set of nodes, $\mathscr{N}$ (also referred to a vetexes), and a set of edges, $\mathscr{E}$, that connect pairs of nodes. For convenience $V = | \mathscr{N} |$ and $E = | \mathscr{E} |$, we chose $V$ (for vertex) to represent the number of nodes to avoid future conflict with $N$ symbolizing finite population size in population genetics.  A node corresponds to a population made up of a large number of random-mating (well mixed) individuals (a Wright-Fisher population, \citep{Fisher1922,Wright1931} with independent Hardy-Weinberg allelic associations, \citep{Hardy1908}) and the edges represent corridors of restricted migration between the populations.  We are also only considering undirected graphs: in the present context this represents equal bidirectional migration between the population nodes.  Furthermore, we are only considering connected graphs (each node can ultimately be visited from every other node) and unlabeled graphs so that isomorphic graphs are considered equivalent. 

The network graph $\mathbb{G}$ is represented by a symmetric $V \times V$ adjacency matrix $\mathscr{A}$.

$$\mathscr{A} =  
\begin{bmatrix}
    a_{1,1} & a_{1,2} & a_{1,3} & \dots  & a_{1,V} \\
    a_{2,1} & a_{2,2} & a_{2,3} & \dots  & a_{2,V} \\
    a_{3,1} & a_{3,2} & a_{3,3} & \dots  & a_{2,V} \\
    \vdots & \vdots & \vdots & \ddots & \vdots \\
    a_{V,1} & a_{V,2} & a_{V,3} & \dots  & a_{V,V}
\end{bmatrix} 
 $$

The presence of an edge between two nodes is represented by a one and the absence of an edge is a zero.  The connectivity of a node is 

$$ c_i = \sum_{j \in \mathscr{N}} a_{i,j}$$

Each generation, $g$, the allele frequency, $p$ of each population node, $i$, is updated with the fraction of immigrants from $n$ adjacent populations, $j$, at a migration rate of $m$.  

$$p_{i,g} = (1 - c_i m) p_{i,g-1} + \sum_{j=1}^N m a_{i,j} p_{j,g-1}$$ 

Note that this equation will not be appropriate if the fraction of alleles introduced into a population exceeded unity. See the discussion of the star topologies illustrated in Figure \ref{fig:simple}.

The frequencies, adjusted for migration, are then paired into genotypes and undergo the effects of  selection.  The remaining allelic transmission sum is normalized by the total transmission of all alleles to the next generation to render an allele frequency from zero to one.  

$$p_{i,g}' = \frac{p_{i,g}^2 + \omega p_{i,g} (1-p_{i,g})}{p_{i,g}^2 + 2 \omega p_{i,g} (1-p_{i,g}) + (1-p_{i,g})^2}$$ 

Note that here for simplicity we set the relative fitness of the homozygotes to one, $w_{AA} = w_{BB} = 1$ and the heterozygote fitness is represented by  $w_{AB} = \omega$.  

\subsection{Analytic Results}

Underdominance in a single population has one central unstable equilibrium and two trivial stable equilibria at $p=0$ and $p=1$.  When one considers multiplying the three fixed points of a single population into multiple dimensions it can be seen that, if migration rates are sufficiently small, nine equilibria ($3 \times 3$) exist in the two-population system. Again, there is a central unstable equilibrium, the two trivial stable equilibria, and two additional internal stable equilibria.  The remaining four fixed points are unstable saddle points that separate the basins of attraction \citep[see Figure 3 of][]{Altrock2010}.  However, as the migration rates increase the internal stable points merge with the neighboring saddle points and become unstable themselves \citep{Karlin1972b,Altrock2010}.  In three populations there are a maximum of 27 ($3 \times 3 \times 3$) fixed points. The three types found in a single population, six internal stable equilibria, and 18 saddle points.  The general pattern is that in $V$ populations there are a maximum of $3^V$ equilibria if migration rates are sufficiently low.  There will always be one central unstable equilibrium (a trajectory starting near this point can end up in any of the basins of attraction) and two trivial fixation or loss points.  A $V$-dimensional hypercube represents the state space of joint allele frequencies of a $V$-population system. This hypercube has $2^V$ vertexes (corners).  It can be seen that the internal stable points, if they exist, are near these corners when the corners represent a mix of zero and one allele frequencies. There are exactly two corners, the trivial global fixation and loss points, that do not contain a mix of unequal coordinate frequencies.  Thus, there are $2^V - 2$ possible internal stable points that represent alternative configurations of a migration-selection equilibrium polymorphism. Finally there are up to $3^V - 2^V - 3$ saddle points that separate $2^V$ possible (but never less than two) domains of attraction. 

We have a system of equations, $p_{i,g}'$, that describe the dynamics of allele frequency change within each population and these dynamics are coupled by migration.  It is useful to think of the difference in frequency each generation, $\delta_{p,i} =p_{i,g}'-p_{i,g-1}'$.
We can set $\delta_{p,\forall i} = 0$ to solve for fixed points in the state space. However, the system needs to be simplified in order to be tractable.  For example if we only look along the $p_1 = p_2$ axis in the two-population case we get three solutions, $p=0$, $p=1$, and $$\hat{p}_{\forall i \in \mathscr{N}}=\frac{w_{AA}-w_{AB}}{w_{AA}+w_{BB}-2 w_{AB}}$$ in the general case and $$\hat{p}_{\forall i \in \mathscr{N}}=\frac{1-\omega}{2-2 \omega} = \frac{1}{2}$$ in the simplified (equal homozygote fitness) case. The first two points are the trivial loss of polymorphism. The third is identical to the internal unstable equilibrium point in a single population \citep{Altrock2010}.

In fact this unstable equilibrium point is always found along the $p_1 = p_2 = \dots = p_V$ axis.  Note that the migration rate, $m$, is not a part of the solution.  The position of this point in the state space is independent of migration rates.  Since it falls along the axis where the allele frequencies of all populations are equal, migration between populations, and in fact the population network topology itself, has no effect.

While the position of this internal point remains fixed regardless of the number of interacting populations in the network, there is a multiple population effect on the rate of change away from this point. Solving for the eigenvalues, $\lambda_i$, of a Jacobian matrix, $\mathbb{J}$, of partial derivatives of the system
along the $p_1 = p_2 = \dots = p_V$ axis shows that the rate of change follows the pattern
$$\left. \lambda_1\right|_{p=\frac{1}{2}} =\frac{2 V}{1+\omega}-V\text{.}$$
Thus, as the number of interconnected populations increase the magnitude of flow away from the internal point increases. 
This rate is a function of both the heterozyote fitness ($\omega$) and the number of interlinked populations. 
At $\omega = 0$, or lethality of the heterozygous condition, the rate of change is equal to the number of connected populations. 

We are also interested in the internal stable equilibria that allow differences in allele frequencies among populations to be maintained. In the two-population special symmetric case this can be solved from $\mathbb{J}$ along the $p_1 = 1 - p_2$ axis and yields $$\lambda_2 = - \frac{1-\omega+2m^*(2m^*-3)}{(1-2m^*)^2} $$ 
$$\left. m^*\right|_{\lambda_2 = 0} = \frac{1}{4} \left( 3 - \sqrt{5+4\omega} \right) $$ \citep[see Appendix A of][for more detail]{Altrock2010}. Unfortunately, with three or more populations, even with highly symmetrical configurations, we have not found a single axis or plane through the state space that captures these internal stable points.  Therefore, we have used numerical methods to characterize the critical points.

\subsection{Numerical Simulations} 

\begin{figure*}
\includegraphics[scale=0.65]{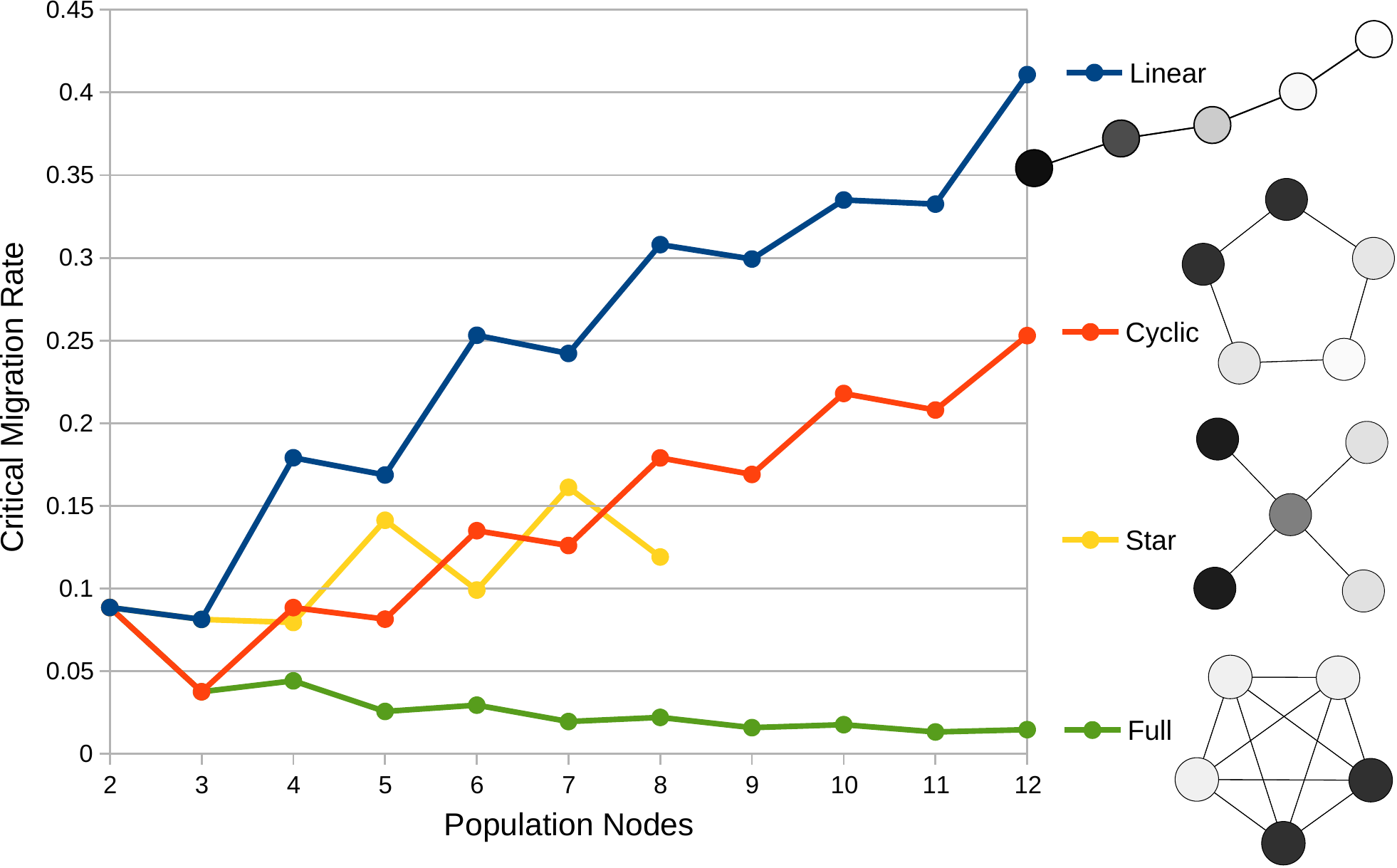}
\caption{\label{fig:simple} 
The stability of some simple network configurations. For each number of nodes and network topology (linear in blue, cyclic in red, star-like in yellow, and fully interconnected in green) the critical migration rate allowing polymorphic underdominant polymorphism to persist at $\omega = 1/2$ is plotted. Examples of each type of network at $V=5$ are plotted as graphs in the legend to the right. The shading of the nodes represents the allele frequency between zero and one of each population near the critical migration rate value.}
\end{figure*}

Sets of populations were initialized with approximately half (depending on if there were an even or odd number of populations) of the nodes in the network at allele frequencies near zero and half near a frequency of one.  The allele frequencies were offset by a small random amount from zero or one to avoid being symmetrically balanced on unstable trajectories.  Migration rates were slowly incremented at each step and the system was allowed to proceed to near equilibrium, when the difference in allele frequencies between generations was less than $10^{-10}$, before the next step. The process was repeated until the point where a collapse in differences of allele frequencies between all populations was detected.  This point was then reported as the critical migration rate of the network.     

\subsubsection{Example Topologies}

The stability of a range of basic network topologies were investigated, shown in Figure \ref{fig:simple}.  In general, in these examples, the diameter of the network is predictive of migration-selection stability.  Linear configurations had the highest stability, cyclic configurations approximately reduced the both the diameter and stability in half.  Fully interconnected populations had both the smallest diameter and lowest stability.  

Another pattern that became apparent is an even-odd alternation in stability.  Except for starlike networks, an odd number of nodes results in a relatively lower stability than an even number of nodes. 

Starlike networks showed an interesting pattern.  They were approximately of the same stability as cyclic networks; however, the even-odd alternation was inverted---odd $V$ graphs showed enhanced stability, showing that the even-odd pattern is not absolute.  At this heterozygote fitness ($\omega = 1/2$) starlike networks with greater than eight nodes could not be evaluated.  At $V\ge 9$ before the critical migration rate is reached the total amount of immigration into the central population exceeds 100\%.  

Fully interconnected networks had the lowest stability and by far the greatest number of edges.  Unlike the other networks the fully-interconnected systems declined at higher $V$.  However, the number of edges grew much faster than the number of nodes. 
Note that the even-odd alternation in relative stability is still apparent, even in these graphs. 

\begin{figure*}
\includegraphics[scale=0.6]{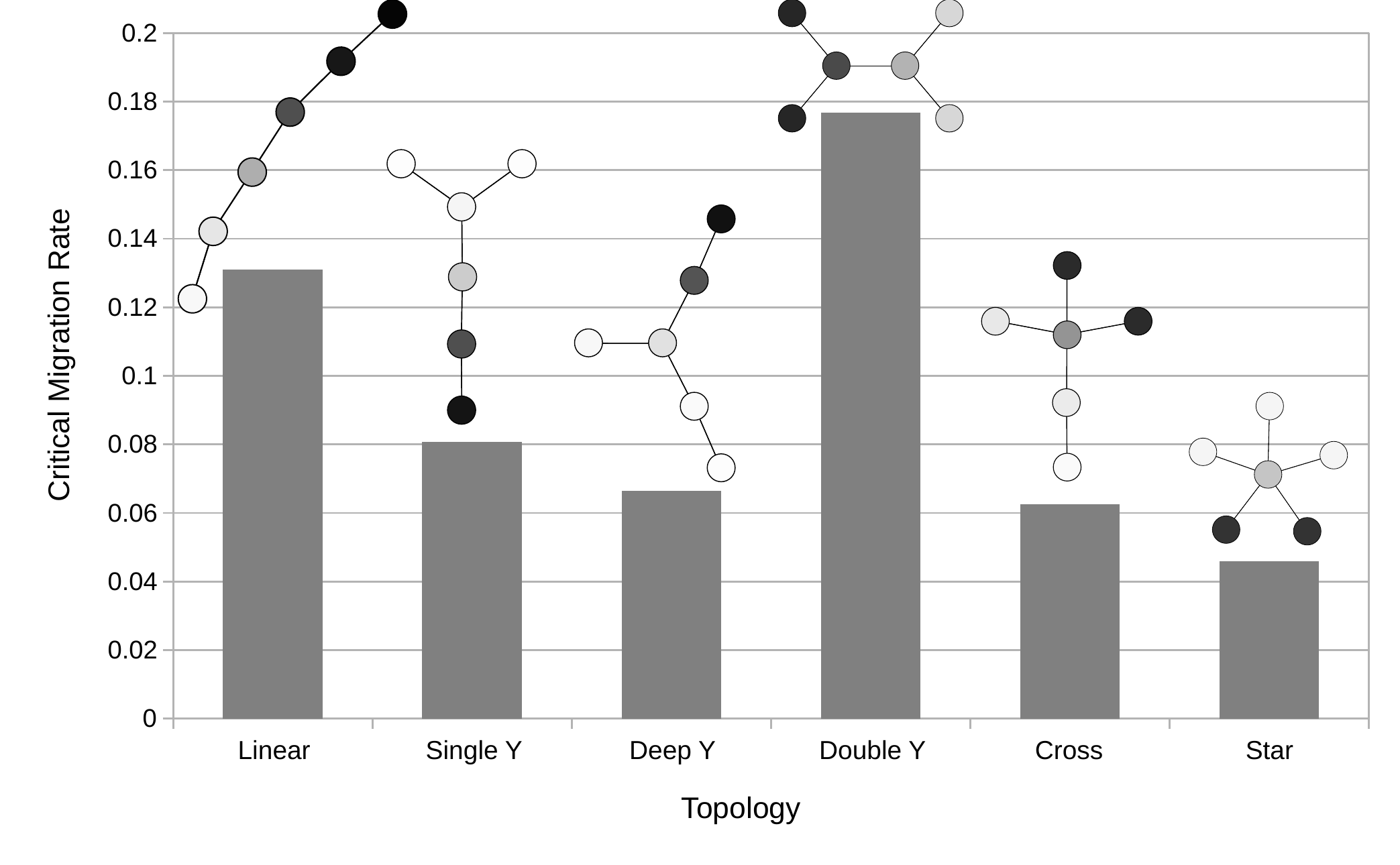}
\caption{\label{fig:6pop} The stability of all possible  simple connected networks made of three populations with five corridors of migration. Here the critical migration rate was evaluated at a relative heterozygote fitness of $\omega = 3/4$. The networks are arranged by diameter of the network declining from five on the left to two on the right.  The shading of the nodes represents the allele frequency between zero and one of each population near the critical migration rate value. The shading of the bars is set to 50\% gray to aid visualization of the allele frequencies of the nodes.}
\end{figure*}

The effects of a range of topologies for six nodes and five edges was also explored, shown in Figure \ref{fig:6pop}.  All of these networks have identical treeness, $\phi_{tree} = 1$ \citep[\textit{sensu}][]{Xie2007}.  In general the stability is correlated with the diameter of the network.  However, the clear exception is the ``double-Y'' topology, which has the highest stability of all.  This has inspired an alternative measure of the treeness of a network that we will refer to as 
``dendricity'' to avoid conflicting with prior definitions of treeness in the literature. 

\clearpage

\clearpage

\subsubsection{Random Graphs}

In order to evaluate general correlations between migration-selection stability and network summary statistics we generated 100,000 random connected graphs of up to 20 nodes in size and evaluated their stability.  The results are summarised in Supplementary Table 1.  We found that the most stable network configurations contained nodes with at most three edges (see the next section below) so, in order to avoid the problem of the rate of immigration more than replacing a local population we set a maximum migration rate of $m = 1/3$ and reported this as $m^*$ for the subset of highly stable network topologies---only 0.64\% of the random networks reached this point of $m=1/3$---these highly stable networks are explored in the next section.  

The following parameters were estimated from these networks: \textit{Variance}, which refers to the variance in connectivity ($c_i$ the number of edges incident with node $i$) of all the nodes in the network. \textit{Efficiency} refers to the shortest path lengths between nodes in the network according to
$$\phi_{efficiency} = \frac{1}{V(V-1)} \sum_{i < j \in V} \frac{1}{d(i,j)} $$
where $d(i,j)$ is the minimum path length between nodes $i$ and $j$. The \textit{diameter} of the graph is the maximum $d(i,j)\in \mathbb{G}$. 
\textit{Dendricity} is the fraction of nodes incident with three edges where at least one edge is a ``bridge'' edge (removing bridge edges results in an unconnected graph) out of the total number of internal nodes.
\textit{Evenness} is simply a binary variable of zero or one to indicate if an odd or even number of nodes are present in the graph.  Finally, \textit{terminalness} indicates the fraction of nodes in the graph that are terminal (or leaf) nodes.  

Each of the summary statistics we addressed were significant predictors of network stability; however, because of correlations between these measures caution must be used to interpret the results.  For example, contrary to intuition the number of nodes was negatively correlated with stability.  This is because the number of possible edges, which generally lower stability, increases dramatically with the number of nodes.  When the number of edges is controlled the number of nodes becomes strongly positively correlated.  Simply the number of nodes per edge ($V/E$) is a powerful predictor of stability.  In general \textit{diameter} is a strong predictor of stability, particularly if the number of nodes are held constant (compare to Figure \ref{fig:6pop}).  \textit{Dendricity} and \textit{terminalness} also performed well as general predictors of stability.  \textit{Evenness} continues to be a predictor of stability, specifically even ordered networks have a greater stability than odd ordered ones, but the predictive power is weak compared to other measures.  
\textit{Variance} and \textit{efficiency} are a bit more difficult to understand.  Increased variance in the number of edges per node is associated with lower stability, while intuitively one might expect the opposite.  When this is measured as \textit{variance} divided by the total number of edges the correlation almost disappears and in fact becomes slightly positive---as expected---unless the number of nodes are held constant. \textit{Efficiency} also varies in a non-intuitive way.  It is positively correlated unless either the number of nodes or the number of edges are held constant where it becomes strongly negatively correlated; however, if both are constant efficiency becomes slightly positively correlated again.  

When exploring model selection to find the minimal adequate model to predict $m^*$ via adjusted $r^2$, 
and  Mallows' $C_p$ as implemented in the \textsf{R} package  ``leaps'' all ten summary statistics were retained using all four methods \citep{R,Lumley2009}.  
Using all of the predictors in the full linear model explains the majority, $r^2=0.72$, of the variation in $m^*$.

\subsubsection{Evolving Networks}

\begin{figure*}
\includegraphics[scale=0.8]{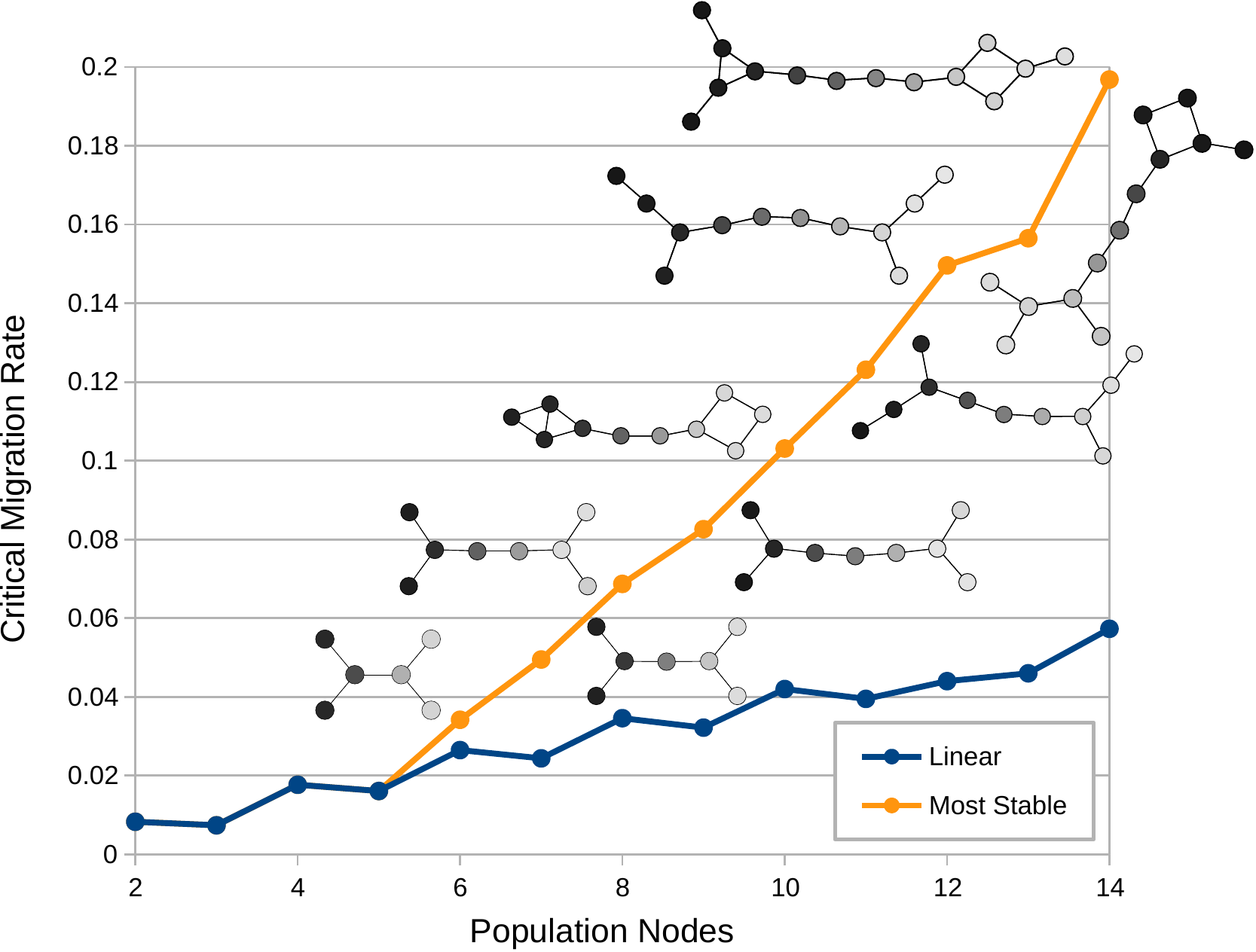}
\caption{\label{fig:evolvestable} A plot of the stability of highly evolved networks ($\omega=0.95$).  For comparison the stability of corresponding linear structures is also plotted (linear, blue). The most stable network found for each $V$ is given (at the end of 10,000 steps of random changes with a Metropolis-Hastings-like chain update, starting from a fully interconnected network, over 8 independent replicate runs), even to the left and odd to the right near each corresponding plotting point. With small change, networks can be substantially more stable than the linear configuration.  This seems to derive from a balance of increasing diameter and forked anchoring structures at either end.}
\end{figure*}

In order to more fully explore the upper edge of highly stable networks for a given $V$ we wrote a program that would make random changes to the network and evolve higher stability configurations.  Starting from a fully interconnected network with as close to half of the nodes near an allele frequency of zero or one as possible, edges were randomly selected to be removed or added with the constraint that the new network remain connected.  Most frequently a single edge was altered but with reducing frequency two or more edges could be changed simultaneously to allow larger jumps in topology and movement away from locally stable configurations. When a network is altered its $m^*_{new}$ value is determined.  If the new critical value is higher than the value of the current network $m^*_{current}$, the new network is adopted for the next step.  If the new critical value is lower than the current network, the new network is adopted with a probability equal to the ratio of the new and current critical migration values ($m^*_{new}/m^*_{current}$).  This also allows the network evolution to explore regions off of local maxima.  Up to $V=5$ the most stable network configuration was a linear topology.  From $V \ge 6$ networks with greater stability than the linear configuration were found and are illustrated in Figure \ref{fig:evolvestable}. Note that the number of possible connected networks increases dramatically with larger $V$. The most stable networks found in Figure \ref{fig:evolvestable} are not expected to result from an exhaustive search, particularly for $V \ge 10$.  They are however meant to illustrate some general properties of highly stable networks.  

\subsection{Software Availability}

All simulations for both random and evolving networks were written in Python 2.7.10. The code is freely available on GitHub:
\url{https://github.com/akijarl/NetworkEvolve} 

\section{\label{Discussion} Discussion}

 One result that is beginning to emerge from the study of evolutionary dynamics on graphs is that the resulting properties can be sensitive to the network topology, but often in a non-intuitive way,  \citep[\textit{e.g.}][]{Hindersin2014,Hindersin2015}.
There are some general factors that influence, or are predictive, of the stability of underdominant polymorphisms in a population network. In smaller networks the influence of diameter and evenness are apparent.  The larger the diameter of the network the effectively lower the migration rate is between the edges of the system, because alleles have to be exchanged via intermediate nodes; it is well understood that a lower migration rate enhances migration-selection stability \citep{Altrock2010}.  In contrast the more edges there are in a system the higher the effective migration rate across the network, which results in lower stability.  

The role of the even-odd number is nodes is more subtle.  In a system of coupled populations having an allele frequency near $p=1/2$ is inherently unstable with underdominant fitness effects.  An odd ordered network that is anchored at high and low frequencies near its edges pushes the central population near $p=1/2$, which destabilizes the entire system. Of course there are exceptions to this rule such as the star network topology  \citep[star networks also have other unusual properties such as acting as amplifiers of selection][]{Frean2013,Adlam2015,Hindersin2015}, and the importance of evenness declines with larger networks.  This pattern may also change if there are three or more alleles that are underdominant with respect to each other.  Interestingly, this effect seems to have been completely overlooked in previous work using either numerical techniques or wave approximations to study underdominant-like effects.  

The networks that were evolved to higher stability illustrate the effects of having two ``anchor'' nodes at each end of a central linear network ``trunk.''  The anchors are made up of variations on a theme of a node with three edges with at least one of these edges being a bridge edge.  Intuitively, the flow of one allele along two paths into a population can overcome the flow of the alternative allele along a single path, thereby enhancing the stability of the anchored edges of the system \citep[\textit{cf.} the discussion of ``stem'' structures, reservoirs, and the movement of clusters of mutants within ``superstar'' networks in][]{JamiesonLane2015}. In contrast a strictly linear network allows adjacent populations to collapse one by one without any local restrictions in gene flow.  Additionally the internal linear trunk has a large diameter further restricting gene flow according to the discussion of the effects of diameter above \citep[compare this to the predictions of underdominance-like dynamics in a continuous population and the tendency of ``tension zones'' to locate in regions of restricted gene flow or population density and stop ``pushed waves''][]{Barton1985, Barton2011}, \citep[see also ``invasion pinning'' in][]{Keitt2001}.

\subsection{Implications}
 
\subsubsection{Natural Systems} 

This work was originally motivated by biological systems.  It is interesting to ask, based on these results, where we might expect to see underdominant polymorphism being maintained in wild populations.   
Population network configurations exist in which even subtle levels of underdomaince can maintain stable geographic differences between populations with substantial rates of migration.  Chromosomal rearrangements  that lead to strong underdominance can occur at relatively high rates and are rapidly established between closely related species \citep{White1978,Jacobs1981}. Subtle underdominant interactions may be more widespread than previously appreciated and may have played a large role in shaping gene regulatory networks \citep{Stewart2013}. Note also that weak effects among loci can essentially self organize to become coupled and that this effect may extend to a broader class of underdominant-like effects such as the well known Dobzhansky-Muller incompatabilites  \citep{Barton2009,Landguth2015}. 
We have found that bifurcating tree-shaped (dendritic) networks have very high stability.  Key natural occurrences of dendritic habitats include freshwater drainage systems, as well as oceanic and terrestrial ridge systems. Indeed, a role of underdominance in shaping patterns of population divergence across connected habitats has been implicated, either directly or indirectly, in freshwater fish species \citep{Fernandes2001,Alves2003,Nolte2009}, salamanders \citep{Fitzpatrick2009,Feist2013}, frogs \citep{Bush1977} (see also \citep{Wilson1974}), semiaquatic marsh rats \citep{Nachman1989}, and \textit{Telmatogeton} flies in Hawai`i,  which rely on freshwater streams for breeding environments \citep{Newman1977} \citep[in the case of Dipterans we are ignoring chromosomal inversions which do not result in underdominance in this group][]{Coyne1991,Coyne1993}.  In fact, alpine valleys around streams also follow a connected treelike branching pattern and there are examples of extensive underdominance in small mammals found in valleys in mountainous regions \citep{Pialek2001,Basset2006,Britton-Davidian2000}.  To the extent that persisting underdominant and underdominant-like fitness effects may promote speciation \citep[rates of karyotype evolution and speciation are correlated][]{Bush1977} it should be noted that freshwater streams contain 40\% of all fish species yet are only 1\% of the available fish habitat and that a higher rate of speciation is indeed inferred for freshwater versus marine systems \citep{Bloom2013}. 

However, there are also examples of the maintenance of underdominant polymorphisms that are not found in species associated with limnological structures. For example, the island of Sulawesi  itself has an unusual branching shape and a large number of terrestrial mammal species with a 90\% or greater rate of endemism excluding bats \citep{Groves2001}.  Other factors that are associated with underdominant stability are the diameter of the network and having an even order of nodes.  The Hawaiian islands essentially form a linear network of four major island groups (Ni`ihau \& Kaua`i---O`ahu---Maui~Nui---Hawai`i) and are known for their high species diversity and rates of speciation with examples in birds \citep{Lerner2011}, spiders \citep{Gillespie2004}, insects \citep{Magnacca2015}, and plants \citep{Helenurm1985,Givnish2009}. For example, the Hawaiian \textit{Drosophila crassifemur} complex has maintained chromosomal rearrangements between the islands that are predicted to result in underdominance \citep{Yoon1975}.  Perhaps the network topology of the Hawaiian archipelago (in addition to the diversity of micro-climates, environments, and ongoing inter-island colonizations) has contributed to the high rates of diversification found on these islands.  

In contrast, areas where we might expect to see less maintenance of genetic diversity that can contribute to boundaries to gene flow are in highly interconnected networks with low diameters such as, perhaps, marine broadcast spawners with long larval survival times that are associated with the shallow waters around islands (\textit{i.e.}, the network nodes).  Examples of a lack of speciation in such groups, distributed over areas as large as half of the Earth's circumference, exist \citep{Palumbi1992,Lessios2003}.  

Another type of network is one that is distributed over time rather than space.  Underdominant interactions have been inferred in the American bellflower \textit{Campanula americana} \citep{Galloway2005}. This species is unusual in that individuals can either be annual or biennial depending on the time of seed germination.   Given that the majority of seeds are expected to germinate within a single year \citep{Galloway2001}, even-year biennials may form a somewhat distinct population from odd-year biennials with gene flow occurring by the subset of annual plants---forming an even ordered network.  Finally, a tantalizing combination exists in the pink salmon, \textit{Oncorhynchus gorbuscha}, of the North Pacific.  This species is both biennial and returns to native freshwater streams to spawn.  Indeed, artificial crosses between even and odd year individuals (of the same species) have revealed extensive genetic differences with hybrid disgenesis \citep{Gharrett1991,Limborg2014}.

\subsubsection{Applications} 

Various ``transgene mitigation'' methods have been proposed to prevent the transfer of genetic modifications from genetically engineered crops to traditional varieties or wild relatives \citep{Lee2006,Daniell2002,Hills2007,Kwit2011} including the use of underdominant constructs \citep{Reeves2014a,Soboleva2003}.  In a species with limited pollen dispersal, it may be tempting to plant a buffer crop area between a GMO crop with underdominant transgene mitigation and an adjacent unmodified population.  However, these results suggest that, over multiple generations, this configuration may actually destabilize the system and promote the spread or loss of the genetic modification.  In a simplistic scenario a single flanking buffer population results in an odd number of populations, breaking the evenness rule of stability (unless the populations are arranged in essentially a $V=5$ star pattern, Figure \ref{fig:simple}). Depending on local conditions, it may be preferable to plant two distinct yet adjacent buffer crops or none at all.  
 Of particular note are genetically modified and wild species that exist in freshwater systems, such as rice \citep{Lu2009} and fish \citep{Devlin2006}. Our predictions suggest that underdominant containment may in general have enhanced stability in these situations. However, this can work both ways.  Genetic modifications may be amiable to underdominant mitigation strategies to prevent establishment in the wild; yet, difficult to remove from freshwater systems if established.

Using underdominance to  stably yet reversibly genetically modify a wild population is one goal within the field of genetic pest management.  The potential implications of these results depend on the amount of modified individuals that could be released into the wild.  If the numbers are sufficiently high to transform an entire region then highly interconnected populations would be ideal to ensure full transformation.  However, as is much more likely, if the number of individuals that can be released is much smaller than the total wild population, transformation might best be achieved in a stepwise strategy utilizing linear or treelike population configurations. In the case of Hawai`i, limiting the effects of avian malaria by modifying non-native Culex mosquitoes has been proposed as a method to prevent further extinctions of native Hawaiian forest birds \citep{Clarke2002}.   Linear island archipelagos and their river valleys \citep[Culex are more common at lower elevations,][]{Van1986} may be ideal cases to both transform local populations yet prevent genetic modifications from becoming established outside of the intended area.  

\subsection{Future Directions} 

It will come as no surprise to an evolutionary biologist that systems with greater genetic isolation (such as freshwater streams versus marine environments) will lead to increased genetic divergence and rates of speciation; however, the implication we are focusing on here is the influence of the population network topology.  We are suggesting that, for the same degree of migration rate isolation, alternative network topologies might be compared to inferred rates of speciation and/or enhanced genetic diversity that leads to hybrid dysgenesis.  A consideration of the geological history is also appropriate to incorporate effects such as stream capture and the merging of islands on biological diversity. A proper meta-analysis or experimental evolution of this network topology effect is beyond the scope of the current manuscript but would be useful projects to further explore these effects.

\section{Competing Interests}
The authors claim no competing interests.

\section{Authors' Contributions}
FAR conceived the study; FAR and \'AJL carried out the programming and analysis; FAR and \'AJL drafted the manuscript. All authors gave final approval for publication. 

\section{Acknowledgments}
We thank Arne Nolte, Mohamed Noor, and Arne Traulsen for helpful comments and Vanessa Reed for copy editing.  

\section{Funding}
This work was supported by a grant  to \'AJL from the Charles H. and Margaret B. Edmondson Fund and a grant to FAR from the Victoria S. and Bradley L. Geist Foundation administered by the Hawai`i Community Foundation Medical Research Program, 12ADVC-51343.

\end{document}